\documentclass[prl, endfloats, superscriptaddress, aps, preprint]{revtex4}
\usepackage{amssymb}
\usepackage{amsmath}
\usepackage{graphicx}
\usepackage{amsfonts}
\usepackage[normalem]{ulem}

\begin{document}
\title{Spatially intertwined superconductivity and charge order in $1T$-TaSSe revealed by scanning tunneling spectroscopy}
\author{Yaroslav A. Gerasimenko$^\dag$}
\email{yaroslav.gerasimenko@ijs.si}
\affiliation{CENN Nanocenter, Jamova 39, SI-1000, Ljubljana, Slovenia}
\affiliation{Department of Complex Matter, Jozef Stefan Institute, Jamova 39, SI-1000, Ljubljana, Slovenia}
\affiliation{Institute of Experimental and Applied Physics, University of Regensburg, Universit{\"a}tstr. 31, 93053 Regensburg, Germany}
\author{Marion A. van Midden$^\dag$}
\affiliation{Department of Solid State Physics, Jozef Stefan Institute, Jamova 39, SI-1000, Ljubljana, Slovenia}
\author{Petra {\v Sutar}}
\affiliation{Department of Complex Matter, Jozef Stefan Institute, Jamova 39, SI-1000, Ljubljana, Slovenia}
\author{Zvonko Jagli{\v c}i{\' c}}
\affiliation{Institute of Mathematics, Physics and Mechanics \& Faculty of Civil and Geodetic Engineering, University of Ljubljana, Jadranska 19, SI-1000, Ljubljana, Slovenia}
\author{Erik Zupani{\v c}}
\affiliation{Department of Solid State Physics, Jozef Stefan Institute, Jamova 39, SI-1000, Ljubljana, Slovenia}
\author{Dragan Mihailovic}
\affiliation{CENN Nanocenter, Jamova 39, SI-1000, Ljubljana, Slovenia}
\affiliation{Department of Complex Matter, Jozef Stefan Institute, Jamova 39, SI-1000, Ljubljana, Slovenia}
\date{\today}

\begin{abstract}
\textbf{
The interplay of different emergent phenomena -- superconductivity (SC) and domain formation -- appearing on different spatial and energy scales are investigated using high-resolution scanning tunneling spectroscopy (STS) in the prototypical transition metal dichalcogenide superconductor $1T$-TaSSe single crystals ($T_{SC} = 3.2$\,K)  at temperatures from 1 to 20\,K. Our major observation is that while the SC gap size smoothly varies on the scale of $\lesssim 10$\,nm, its spatial distribution is not correlated to the domain structure. On the other hand, there is statistically significant correlation of the SC gap $\Delta_{SC}$ with spectral weight of the narrow band at the Fermi level formed from the same Ta $5d$ orbitals as the Mott-Hubbard band. We show that the narrow band follows the evolution of Hubbard bands in space, proving unambiguously its relation to the charge order. The correlations between the two suggest a non-trivial link between rapidly spatially varying charge order and superconductivity common in many quantum materials, and high-temperature superconductors in particular.}
\end{abstract}

\maketitle
\def\thefootnote{\dag}\footnotetext{These authors contributed equally to this work}\def\thefootnote{\arabic{footnote}}

\section*{Introduction}
The problem of how the interplay of interactions on multiple energy, spatial and time scales lead to the formation of emergent quantum matter\cite{Basov17} touches upon many areas, ranging from unconventional superconductivity\cite{FKT15} to electronic stripes\cite{Tranquada95}, Wigner crystals\cite{Shapir19, Madhavan19} and colossal magnetoresistance\cite{Dagotto05}. Self-organized spatial inhomogeneity has been a topic of interest, as well as controversy, in the study of superconductivity in high-temperature cuprate superconductors ever since their discovery, where scanning tunneling microscopy studies give valuable insights\cite{Fisher,Pan01}. 
In layered transition metal dichalcogenides (TMDs), various charge orders emerge from the intricate interplay of electron-electron, $(e\text{-}e)$,\cite{Sipos08,Kogar17,Vodeb19} and electron-phonon\cite{Rossnagel11,Butler21} interactions. The two are relatively close in energy\cite{Rossnagel11} and exhibit complex competition on various spatial scales\cite{Kim94,Cho16,Ma16}, but recently were nicely separated in time domain\cite{Petersen11,Hellmann12,Porer14}. Such complexity gives rise to rich phase diagrams\cite{Sipos08,Morosan06,Wang18,Ribak20} that can be even further extended in the time domain to metastable charge configurations created with ultrashort light pulses\cite{Stojchevska14,Han15,Zong18,Gerasimenko19hidden,Ravnik19,Ravnik20,Stahl20,Ravnik21qb}.

\begin{figure}[p]
\includegraphics[width=0.8\columnwidth]{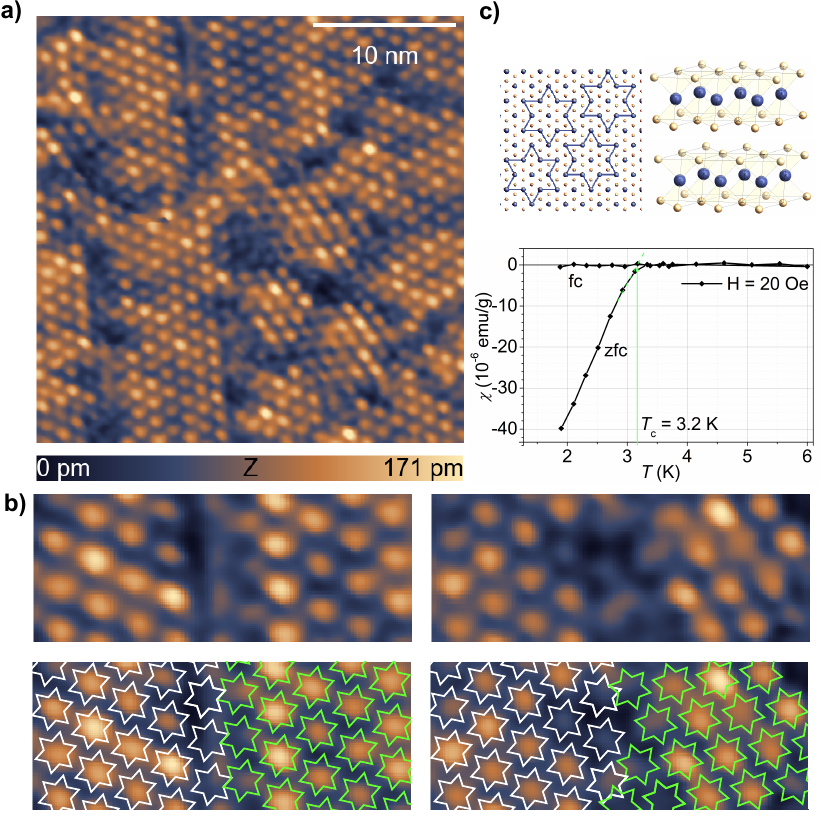}
\caption{\textbf{Charge-density wave and domain structures in 1T-TaSSe:}
(a) Characteristic STM topography ($V = -50$\,meV, $I = 1.5$\,nA) and (b) zoom into phase and chiral domain walls with corresponding star-of-David assignment. Note the shift and the rotation between CDW domains separated by phase (left) and chiral (right) domain walls respectively.
(c) Top: crystal structure of $1T$-TaSSe (Ta -- blue, S, Se -- yellow), David star pattern show the $\sqrt13\times\sqrt13$ CDW reconstruction; bottom: temperature dependence of magnetic susceptibility, $\chi(T)$, showing the superconducting transition in the studied bulk crystal (see Methods section).
}
\label{fig:topo}
\end{figure}

Most remarkably, the superconductivity (SC) in the most well-studied TMDs, $1T$-TaS$_2$\cite{Sipos08, Li12, Liu13, Ritschel13,Yu15gate,Qiao17} and $1T$-TiSe$_2$\cite{Morosan06,Kusmartseva09,Joe14}, is \emph{universally} correlated with the formation of domains in the charge order, separated by phase-shifted domain walls (Fig.~\ref{fig:topo}a,b).
 
A very natural scenario for the formation of intertwined orders\cite{FKT15}, suggests that superconductivity can emerge within gapless topological defects of a symmetry-broken state\cite{GorkovLebed83,KivelsonEmery96}, e.g. in the domain walls of an electron-phonon driven charge-density wave. 
Indeed, evidence towards intimate relation between SC and DWs was recently demonstrated in electrostatically doped $1T$-TiSe$_2$\cite{Li16LP}, where the observation of Little-Parks oscillations was interpreted in favor of superconducting domain walls coexisting with insulating charge-ordered domains. However, recent studies\cite{Cho17walls,Skolimowski19,Gao19PG} have challenged this hypothesis in $1T$-TaS$_2$, where isolated domain walls were observed to remain gapped due to $e\text{-}e$ interactions, suggesting the above simple picture is inherently incomplete. The crucial and universal question that has implications beyond superconductivity in TMDs is thus what part of charge order does superconductivity couple to and, hence, whether domain walls are necessary or just a mere coincidence?

In its parent compound, $1T$-TaS$_2$, interplay between Coulomb interaction, lattice strain and Fermi surface instability leads to a complex phase diagram, where in the ground state a $\sqrt{13}\times\sqrt{13}$ CDW\cite{Rossnagel11} coexists with a Mott insulator (MI)\cite{FazekasTosatti,Sipos08} and quantum spin liquid\cite{Klanjsek17,LawLee17}. Upon isovalent substitution of Se for S (``chemical pressure''), the uniform CDW is suppressed in favour of the domain state, concomitant with the simultaneous Mott gap collapse\cite{Ang13SeS} and insulator-to-metal transition\cite{Liu13}. Superconductivity emerges next to the MI endpoint and spans a small range of Se content, vanishing towards $1T$-TaSe$_2$\cite{Liu13}. Photoemission studies show that SC emerges from the same Ta $5d$ carriers as CDW and MI, rather than from Se $4p$, and thus suggests that they coexist in real space\cite{Ang13SeS}. More detailed studies under pressure also reveal the clear separation between the uniform CDW + MI suppression and SC onset on the phase diagram\cite{Sipos08,Ritschel13}, suggesting that the details of the domain-like charge order are important.

Microscopic studies of the superconductive $1T$-TaS$_{2-x}$Se$_x$ at high temperatures, $T>T_{SC}$, were able to link Mottness collapse and metallicity to the Se-induced buckling of the David star cluster\cite{Qiao17}, apparently related to the Se ordering within it\cite{Ang15}. However the relation of these very local nm-scale changes to superconductivity remains unclear. Furthermore, the degree of electronic disorder depends on the energy scale: it increases from high, of the order of the MI gap, to low ones, of the order of the expected SC gap, apparently disentangling the two phenomena and suggesting superconductivity within an emergent honeycomb network\cite{Gao19PG}. However, so far, no experimental spatial mapping of the SC gap has been done in these compounds that could directly and unambiguously elucidate the link between SC and charge order. 

Here we study the superconductive $1T$-TaSSe single crystal with a nearly optimal $T_{SC} = 3.2$\,K using high-resolution scanning tunneling spectroscopy at temperatures from 1 to 20\,K. Our major observation is that while the SC gap size smoothly varies on the scale of $\lesssim 10$\,nm, its spatial distribution is not correlated to the domain structure. On the other hand, there is statistically significant correlation of the gap size with the fine tuning of the position of the narrow band in relation to the Fermi level. We show that the band follows the evolution of Hubbard bands in space, providing a direct link to the MI charge order. The spatial distribution is unrelated to the structural features and is likely linked to the local changes to the David star cluster induced by the Se content. The correlations between the two suggest a non-trivial link between the charge order and superconductivity.

\section*{Results}
To study the possible correlations between CDW and SC orders, we first compare the characteristic length-scales over which CDW and SC order parameters change. In a CDW, the charge modulation and lattice distortion are inseparable and thus the former can be easily seen in the topographic images. Figure~\ref{fig:topo}a shows the complex CDW domain structure of $1T$-TaSSe. Inside the domains, the CDW is structurally commensurate with the atomic lattice (Fig.~\ref{fig:topo}c), and the order parameter is constant\cite{NakanishiShibaNC}. The domain structure is therefore characterized by the 26 possible discrete phase shifts of the commensurate CDW\cite{Huang17,Karpov18}. The shifts occur within domain walls -- either translational, or chiral -- that can be seen as either a translation or rotation of the CDW order in the domains they separate (Fig.~\ref{fig:topo}b,c). The order parameter is suppressed inside the domain walls, but does not vanish. Thus, the characteristic scale is set by the domain size $\xi_{CDW}\approx6.8$\,nm (for the details of definition, see SI). The above value is consistent with the previous estimates from X-ray diffraction studies of superconducting $1T$-TaS$_2$ under pressure\cite{Ritschel13}.

\begin{figure}[p]
\includegraphics[width=0.65\columnwidth]{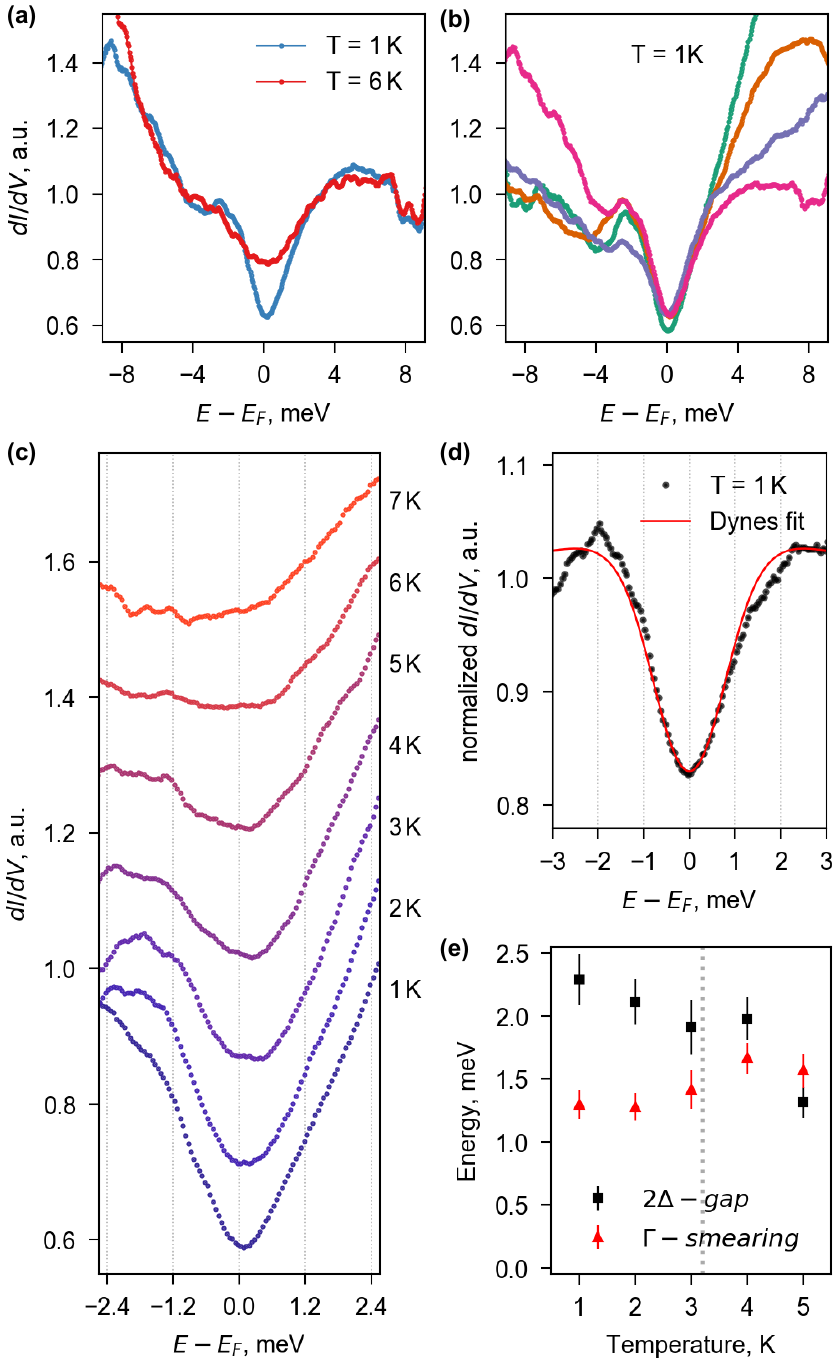}
\caption{\textbf{Temperature evolution of the SC gap:}
(a) Tunneling spectra below and above $T_{SC}$ (gap preset: $V=30$\,mV, $I=1.2$\,nA);
(b) Variation of characteristic tunneling spectra taken on the $2\times2$\,nm$^2$ area at 1\,K;
(c) Evolution of the tunneling spectra taken in the same spot of the sample with temperature, the curves are offset for clarity;
(d) Dynes fit of the $dI/dV$ curve at 1\,K normalized by that at 7\,K;
(e) SC gap and smearing extracted from Dynes fit of the spectra in (b) normalized by the 7\,K curve.
}
\label{fig:tdep}
\end{figure}

The SC gap can be reliably identified despite the complexity of local spectra as shown in Figure~\ref{fig:tdep}. The dip of $2\Delta_{SC}\approx2.2$\,meV width, and centered at $E_F$, disappears upon raising temperature from $T=1$\,K to $T=6\,\mathrm{K}>T_{SC}$ that allows us to associate it with the SC transition (Fig.~\ref{fig:tdep}a). We readily see that the SC gap exists on top of very different backgrounds (Fig.~\ref{fig:tdep}b), signifying that the local structure is important.  
While certain background spectra might be argued to represent some kind of pseudogap, strong asymmetry, detuning from $E_F$ and lack of temperature dependence allows us to rule out any other gaps besides $\Delta_{SC}$ in the $\pm10$\,meV range.

Figure~\ref{fig:tdep}b shows the temperature-dependent spectra measured in the same spot of the sample (see SI for the STS energy resolution). The dip gradually disappears with increasing the temperature (cf. Fig.~\ref{fig:tdep}a,b), until the spectra at 6 and 7\,K look almost identical. Thus the local transition temperature $5\,\mathrm{K}<T_{SC}^{loc}\lesssim6$\,K is substantially larger than the macroscopic $T_{SC}\approx3.2$\,K. Next, we normalize the lower temperature spectra by the 7\,K spectrum and fit them with Dynes formula\cite{Dynes78} to extract the temperature dependence of the gap size $2\Delta_{SC}$ and the incoherent smearing $\Gamma$ (Fig.~\ref{fig:tdep}c). SC suppression in the accessible temperature range is associated mostly with the reduction of the gap size, rather than with scattering, unlike in the conventional SC\cite{Dynes78}. The local ratio $\Delta_{SC}/T_{SC}^{{loc}}\approx2$ is close to the mean-field value of 1.76.

\begin{figure*}[t!]
\includegraphics[width=0.99\textwidth]{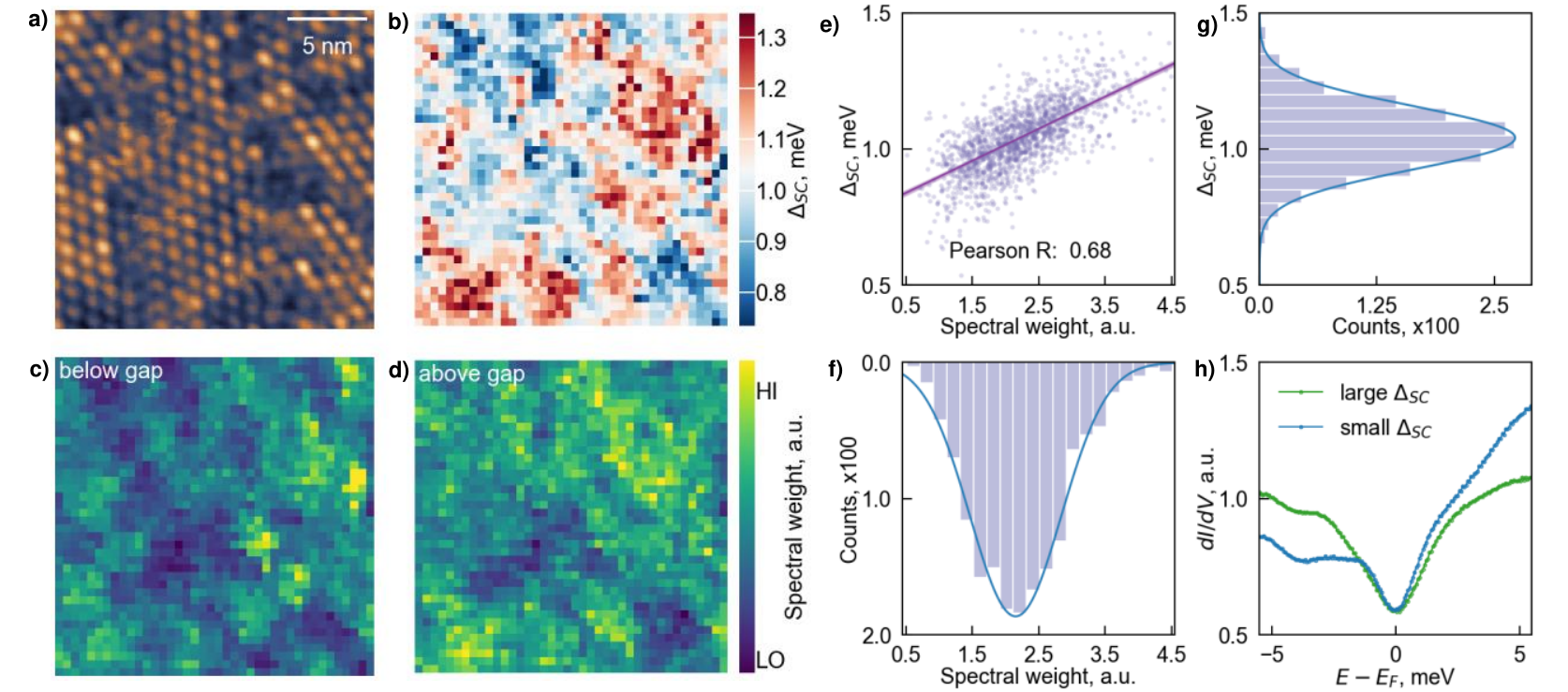}
\caption{\textbf{Correlating SC gap to structural and spectral features:}
(a) Topographic STM image ($20\times20$\,nm, $V=-50\,\mathrm{mV},\,I=1.5\,\mathrm{nA}$) and the corresponding spatial distribution of the SC gap size (b) extracted from the STS map (gap preset: $V=-50$\,mV, $I=2$\,nA);
(c, d) Spatial map of the spectral weight below and above the SC gap obtained by integrating the density of states over $[-10, -2]$\,meV and $[2, 10]$\,meV intervals respectively.
(e) Co-plot of the SC gap $vs$ the spectral weight and the linear regression between the two (solid line; shaded area shows error bar range).
(f, g) Histograms of the spectral weight and SC gap values plotted in (e). Solid lines show the Gaussian fits.
(h) Averaged tunneling spectra in the areas of large and small SC gaps in (b).
}
\label{fig:mapreg}
\end{figure*}

Side-by-side comparison of the spatial distribution of the SC gap size and CDW domain structure in Figure~\ref{fig:mapreg}a,b readily shows that the they are not correlated. Three things are apparent from the spatial distribution of the SC gap: (i) it does not vanish anywhere, (ii) it smoothly varies in space on the $\sim10$\,nm scale (distance between dark-red and dark-blue parts in Fig.~\ref{fig:mapreg}b), (iii) it has local several nm-scale fine structure with rather small ($\sim5$\%) changes in the gap size.  A coarse large-scale mapping of the SC gap (see SI) gives more precise ``correlation'' length $\xi_{SC}\approx 9.3$\,nm. Spatial scale of the gap size variation is known to correspond reasonably well to the Ginzburg-Landau correlation length\cite{Sacepe08}, and indeed $\xi_{SC}$ value is in a reasonable agreement with the prior measurements of $H_{c2}$ in the electrostatically doped 1T-TaS$_2$\cite{Yu15gate} (see SI for discussion). Comparison of $\xi_{SC}$ \emph{vs} $\xi_{CDW}$ shows quantitatively that there is no clear real-space correlation between SC and CDW.

At the same time, local variations of the SC gap size can be correlated with the distribution of the \textit{spectral weight} (SW) in the vicinity of Fermi level. The low-energy background spectrum at $T>T_{SC}$ (Fig.~\ref{fig:tdep}a) changes on small spatial scales. The integrated $\pm10$\,meV SW outside the SC gap (Fig.~\ref{fig:mapreg}c,d) varies on similar scales. A plot of the SC gap vs. SW in each point of the map shown in Fig.~\ref{fig:mapreg}e vividly demonstrates a correlation between the two. The Gaussian distribution of both quantities allows us to use Pearson criterion that gives significant correlation of $R=0.68$. A slight asymmetry in the distribution of SW towards higher values that is absent in the SC gap is apparently due to large $\xi_{SC}$ that keeps the latter finite everywhere. SC gap size is equally well correlated with SW below and above $E_F$ in the available energy window, suggesting that it is the total density of states that is relevant rather than a fine tuning of a certain energy level with respect to $E_F$.

\begin{figure}[hbtp]
\includegraphics[width=0.75\columnwidth]{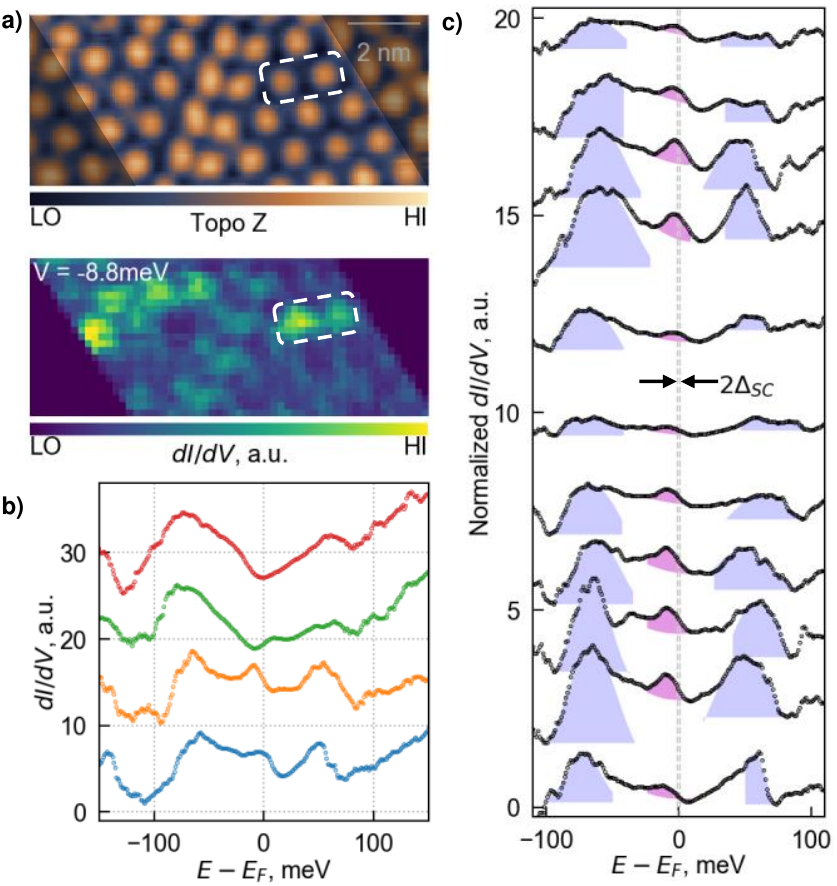}
\caption{\textbf{The origin of the midgap spectral weight:}
(a) Topographic map ($V=700$\,mV, $I=100$\,pA) and density of states at 20\,K (gap preset: $2.2\mathrm{\r{A}}$ approach relative to $V=480$\,mV, $I=100$\,pA )
(b) characteristic spectra (the curves are offset for clarity)
(c) evolution of the midgap spectral weight between two neighboring David stars (marked with dashed rectangle in (a); the curves are offset for clarity). Upper and lower Hubbard bands are shaded with light-blue, the midgap states -- with magenta.
}
\label{fig:midgap}
\end{figure}

Tunneling spectroscopy over a larger energy window shown in Figure~\ref{fig:midgap} allows us to link the low-energy SW to the small band emerging as a result of correlated charge order. Spectra measured at $T=20$\,K, well above $T_{SC}$, or $T_{SC}^{loc}$ (Fig.~\ref{fig:midgap}b), all show  a finite density of states at $E_F$ with two large peaks $\sim150$\,meV apart, lower and upper Hubbard bands, -- characteristic of a collapsed Mott state. On the other hand, the low-energy region reveals a variety of behaviours ranging from a local minimum to a 20\,meV-wide hump moving in the vicinity of $E_F$ (Fig.~\ref{fig:midgap}b). The hump is a very local feature, changing on the scale of a single David star cluster and is unrelated to the domain wall (Fig.~\ref{fig:midgap}a). In Figure~\ref{fig:midgap}c we follow the spatial evolution of the spectra within the one such feature along the long edge of the marked dashed rectangle. It reveals that the hump nicely follows the evolution of intensity of the Hubbard bands, reaching maximal strength at the center of the David stars and vanishing in between, -- the ``smoking gun'' evidence of its relation to Mott localization\cite{Kim94}. This is the unambiguous evidence that the hump, responsible for the low-energy SW promoting SC pairing, originates from the same Ta $5d$ electrons as the charge Mott-localized in the center of the David star. This is also in line with the ARPES data suggesting that SC stems from the Ta orbitals\cite{Ang13SeS}.

The spatial distribution of the low-energy SW is remarkably independent from the nearest energy scale, the Hubbard bands. The latter forms regular lattices\cite{Qiao17,Gao19PG}  and slightly move with respect to each other. On the other hand, SW and hump distributions are significantly more inhomogeneous -- a feature also attributed to mutlifractality\cite{Gao19PG}. 

\section*{Discussion}

Intertwined orders derive its distinctive features from the ``topological doping''\cite{KivelsonEmery96} physics of 1D systems, but the picture changes substantially in 2D. Upon doping a 1D system, it is energetically favourable to locally trap charges within defects of the order parameter topological (solitons), rather than let them populate the conduction band. For a discrete symmetry 1D system like polyacetylene, order parameter is necessarily zero inside the soliton, hence the density wave gap is closed forming a ``metalic'' state\cite{BGS82}. Such a metallic/superconducting state intertwined with a spin-density wave apparently exists in quasi-1D Bechgaard salts close to the SDW endpoint \cite{GorkovLebed83,Kang10,Gerasimenko14}. However, the situation is different in 2D systems like $1T$-TaS$_2$, where Landau theory of CDW transitions \cite{McMillan76,NakanishiShibaNC} suggests that the CDW order parameter is only partially suppressed inside the domain wall. Indeed, conduction and valence band peaks visible in tunneling spectroscopy do not merge inside the domain walls\cite{Cho17walls,Skolimowski19}, leaving a gap open in the DOS. This suggests an explanation for the negligible link between the CDW domain structure and superconductivity in $1T$-TaS$_2$-based materials.

In general, the influence of Coulomb interactions on the topological walls in density wave insulators is still an open question. For single-standing domain walls, it was shown to be enough to cause Mottness collapse, but that the formation of the expected metallic quasiparticle band is suppressed by various competing processes like bond dimerization\cite{Skolimowski19}, band reconstruction\cite{Cho17walls} or a proposed 1D Wigner crystal formation\cite{Aishwarya19}. More complex picture emerges in the domain wall networks\cite{Ma16,Cho16,Park19}, but the characteristic suppression of the Fermi level density of states is still present inside the domain walls, suggesting that domain walls are still a bad host for superconductivity.

Se substitution affects the correlated states differently, increasing the SW at the Fermi level and promoting SC. Possible origin for such behavior could stem from the David star buckling caused by larger size of Se compared to S, as was suggested recently in DFT+U simulations\cite{Qiao17}. We indeed observe that Hubbard bands move closer when the low-energy hump is present, which could suggest that the Mott state is collapsed by the distortions caused by Se and the hump is related to the quasiparticle peak\cite{Qiao17}. Comparing it to the effect of domain walls we readily notice that the above reasons for preempting the quasiparticle band are absent in such local picture. The role of chalcogen atoms is also widely discussed in the 3D stacking and its relation to the metal-to-insulator transition\cite{Lee19}, implying that interlayer correlations might also play its role in the Mott state suppression\cite{Butler19}. 

We conclude that the main mechanism by which SC is induced in $1T$-TaS$_2$-based materials is through pulling states from Mott states to a narrow band at the Fermi level, which supports pairing and phase coherence over spatial extent greater than the characteristic spatial scale of the domain structure.
These results are consistent with in Cu-intercalated $1T$-TiSe$_{2}$ where there is a shift of the Fermi level was suggested to be caused by aggregated Cu\cite{Spera19}. This discussion of spatial distribution of the pseudogap and SC gap is highly relevant to high-temperature superconducting cuprates, where a similarly uniform SC gap was found to coexist with short $1\sim2$ nm length-scale variations of the pseudogap features\cite{Fisher, Pan01}.

Finally, we would like to comment on the possible SC in metastable states in $1T$-TaS$_2$ and similar materials, where domain structures are created with optical pulses\cite{Stojchevska14, Gerasimenko19hidden}. It may be surprising that none of these states have been reported to become superconducting so far, but to our knowledge metastable SC at sufficiently low temperatures has not yet been investigated. But the relevant spectral signatures in confined structures\cite{Ravnik21qb} that are present already at high temperatures reveal intra-gap states that are highly suggestive that SC may be revealed by appropriate low-temperature measurement. Moreover, we were able to confirm that SC persists in the jammed amorphous state\cite{Gerasimenko19j,Vodeb19} in $1T$-TaSSe (see SI), making its analog created by ultrafast optical or electrical pulses a favorable candidate for metastable SC.

\section*{Methods}
\textbf{Sample growth and characterization}
$1T$-TaSSe single crystals were synthesized  with the iodine-assisted physical vapor transfer technique. Crystal structure and surface composition were verified with single-crystal X-ray diffraction measurements and energy-dispersion spectroscopy respectively. Bulk superconducting transition temperature was determined as the onset of diamagnetic response in temperature dependence of susceptibility, $\chi(T)$, measured with QuantumDesign MPMS-7 system. Measurements were done on the single crystal of the mass of $25.15$\,mg with the field of $20$\,Oe applied in the plane of the layers.

\textbf{STM sample preparation and measurements}
The experiments were performed using the commercial SPECS Joule-Thompson STM operating at the base temperature of 1.2\,K and the base pressure below $10^{-10}$ mbar. $1T$-TaSSe single crystals were mounted on a sample holder with a conductive epoxy and cleaved in situ with Kapton tape inside a UHV chamber at room temperature. Samples were immediately transferred to the STM head afterwards and gradually cooled down to the base temperature.

The measurements were performed with either etched W or cur Pt/Ir tips verified on Cu(111) surface. Spectroscopic resolution was calibrated against the superconducting gap in Pb(111) single crystal (see SI for details). Topographic images were taken in the constant current mode. Grid spectroscopic measurements were performed using the standard lock-in technique with AC modulation down to 150\,uV at the frequency of 733\,Hz. For SC spectra measurements, the tip was first stabilized at small setpoint current and bias (values) and then approached closer to the sample with the open feedback loop to increase current and signal-to-noise ratio. The parameters were chosen based on the large energy scale features in the studied area.


\bibliography{stm-doped_arxiv}
\bibliographystyle{apsrev4-1}
\newpage
\end{document}